\documentclass[12pt]{article}
\usepackage[english]{babel}
\usepackage[latin1]{inputenc}
\pdfoutput=1

\usepackage{epsfig}

\usepackage{color}
\usepackage{amssymb}
\usepackage{graphicx}
\usepackage{slashed}
\usepackage[text={16.4cm,23.7cm}]{geometry}

\usepackage{epsfig,psfrag,graphics,verbatim}
\usepackage{epstopdf}
\usepackage{dcolumn}
\usepackage{bm}
\usepackage{amsmath}
\usepackage{float}
\usepackage{placeins}

\newcommand{\D}{\text{d}}
\newcommand{\be}{\begin{equation}}
\newcommand{\ee}{\end{equation}}
\newcommand{\bea}{\begin{eqnarray}}
\newcommand{\eea}{\end{eqnarray}}

\begin{document}

\begin{flushright}
\small KIAS-P13012\\
FLAVOUR(267104)-ERC-37\\
TUM-HEP 879/13
\end{flushright}

\vspace{1cm}
\begin{center}
{\huge\bf  Gamma-ray boxes \\[3mm] from axion-mediated dark matter}\\
\bigskip \vspace{1.5cm}{
{\large\bf Alejandro Ibarra$^a$, Hyun Min Lee$^b$, Sergio L\'opez Gehler$^{a,c}$, \\ \vspace{0.1cm} Wan-Il Park$^d$ and Miguel Pato$^e$}
\vspace{0.5cm}
} \\[7mm]

{\em $^a$Physik-Department T30d, Technische Universit\"at M\"unchen, \\ James-Franck-Stra\ss{}e, D-85748 Garching, Germany}\\
{\it $^b$Department of Physics, Chung-Ang University, 06974 Seoul, Korea}\\
{\it $^c$Excellence Cluster Universe, Technische Universit\"at M\"unchen, \\ Boltzmannstra\ss{}e 2, D-85748, Garching, Germany}\\
{\it $^d$Department de F\'isica Te\`orica and IFIC, \\ Universitat de Val\`encia-CSIC, E-46100, Burjassot, Spain}\\
{\it $^e$The Oskar Klein Centre for Cosmoparticle Physics, Department of Physics, \\ Stockholm University, AlbaNova, SE-106 91 Stockholm, Sweden}\\

\end{center}
\bigskip
\begin{abstract}
We compute the gamma-ray output of axion-mediated dark matter and derive the corresponding constraints set by recent data. In such scenarios the dark matter candidate is a Dirac fermion that pair-annihilates into axions and/or scalars. Provided that the axion decays (at least partly) into photons, these models naturally give rise to a box-shaped gamma-ray spectrum that may present two distinct phenomenological behaviours: a narrow box, resembling a line at half the dark matter mass, or a wide box, spanning an extensive energy range up to the dark matter mass. Remarkably, we find that in both cases a sizable gamma-ray flux is predicted for a thermal relic without fine-tuning the model parameters nor invoking boost factors. This large output is in line with recent Fermi-LAT observations towards the Galactic centre region and is on the verge of being excluded. We then make use of the Fermi-LAT and H.E.S.S.~data to derive robust, model-independent upper limits on the dark matter annihilation cross section for the narrow and wide box scenarios. H.E.S.S.~constraints, in particular, turn out to match the ones from Fermi-LAT at hundreds of GeV and extend to multi-TeV masses. Future Cherenkov telescopes will likely probe gamma-ray boxes from thermal dark matter relics in the whole multi-TeV range, a region hardly accessible to direct detection, collider searches and other indirect detection strategies.
\end{abstract}

\thispagestyle{empty}

\normalsize

\newpage

\setcounter{page}{1}

\section{Introduction}

\par There is mounting evidence for the existence of dark matter (DM) in our universe through its gravitational interactions with ordinary matter (see for instance \cite{Bertone:2004pz,Bergstrom:2012fi} for reviews). However, there is currently no unequivocal indication for non-gravitational interactions neither in direct, indirect nor collider searches. Good prospects exists, though, if the dark matter particle is a thermal relic with a mass of the order of the electroweak symmetry breaking scale \cite{Bertone:2010at}. In this case, the interaction rate with nuclei, the annihilation rate in galaxies or the production rate at colliders could be large enough to allow detection, provided the theoretical and experimental backgrounds are sufficiently well-understood and suppressed.

\par A promising strategy to detect dark matter particles consists in the search for gamma-ray spectral features produced in dark matter annihilations. Such features could stand out over the featureless energy spectrum of the diffuse gamma-ray background or the known astrophysical sources even for moderately small annihilation rates. Up to now three sharp spectral features have been identified: gamma-ray lines \cite{Srednicki:1985sf,Rudaz:1986db,Bergstrom:1988fp}, internal bremsstrahlung \cite{Bergstrom:1989jr,Flores:1989ru,Bringmann:2007nk} and gamma-ray boxes \cite{Ibarra:2012dw}. Gamma-ray lines appear in the annihilation or the decay of dark matter particles into two daughter particles, one of which is a photon, hence producing a spectrum with a monoenergetic line. Internal bremsstrahlung occurs, for example, in annihilations of Majorana or scalar dark matter particles into a photon plus a relativistic fermion-antifermion pair, giving rise to an energy spectrum with a prominent feature close to the endpoint. Lastly, a gamma-ray box arises in the annihilation or decay of dark matter particles into a pair of intermediate scalars, which in turn decay into two particles, one of them a photon. Then, the photon, which is monoenergetic in the rest frame of the intermediate particle, is boosted in the Galactic frame and acquires an energy that depends on the angle between the momentum of the photon and the momentum of the intermediate scalar, resulting in a box-shaped spectrum. 

\par Several searches for these gamma-ray features have been conducted in recent years. Gamma-ray lines have been searched for in the Galactic centre \cite{Vertongen:2011mu,Ackermann:2012qk,Abramowski:2013ax}, in the isotropic diffuse gamma-ray background \cite{Abdo:2010dk,Abramowski:2013ax} and in dwarf galaxies \cite{Aleksic:2012cp}, probing at present the range of dark matter masses between 1 GeV and 20 TeV. Besides, signatures from internal bremsstrahlung have been searched for in the Galactic centre \cite{Bringmann:2012vr,Abramowski:2013ax} and in the isotropic diffuse gamma-ray background \cite{Abramowski:2013ax}, covering the mass range $40\,{\rm GeV}-300\,{\rm GeV}$ and $500\,{\rm GeV}-20\,{\rm TeV}$. Finally, gamma-ray boxes have been searched for in the Galactic centre and halo regions \cite{Ibarra:2012dw} for dark matter masses in the range $5\,{\rm GeV}-800\,{\rm GeV}$. No sign of dark matter annihilations has been unequivocally found, although an intriguing hint for a sharp feature at $\sim 130$ GeV has been reported in \cite{Bringmann:2012vr,Weniger:2012tx,Su:2012ft}. It is at the moment unclear whether this feature is indeed the result of dark matter annihilations or, on the contrary, is an instrumental artifact or a statistical fluke. Future observatories have, however, good prospects to ellucidate whether this gamma-ray excess is genuine or not \cite{Bergstrom:2012vd}.

\par From the theoretical side, previous works have discussed the phenomenology of particle physics models which generate gamma-ray lines \cite{Jungman:1994cg,Bergstrom:1997fh,Bern:1997ng,Bergstrom:1997fj,Gustafsson:2007pc,Bertone:2009cb,Dudas:2009uq,Mambrini:2009ad,Jackson:2009kg,Chu:2012qy}, signals from internal bremsstrahlung \cite{Bell:2008ey,Barger:2009xe,Garny:2011cj,Garny:2011ii,Asano:2011ik} or gamma-ray boxes \cite{Ibarra:2012dw,Chu:2012qy,Fan:2012gr}. In this paper we shall focus on the axion-mediated dark matter model recently proposed in Refs.~\cite{axion1,axion15,axion2} and discuss a specific realisation that generates a box-shaped gamma-ray spectrum. This spectrum falls naturally into one of two possible categories: a narrow box or a wide box. In both cases thermal relics produce gamma-ray outputs at the reach of current experiments. The search for gamma-ray boxes is thus well-motivated from the theoretical viewpoint and, on that respect, the next generation of very high-energy gamma-ray telescopes (such as CTA \cite{Consortium:2010bc}) is eagerly awaited.

\par The paper is organised as follows. In Section \ref{sec:axion} we introduce the main features of the axion-mediated dark matter model and in Section \ref{sec:annihilations} all the relevant tree-level annihilation cross sections are explicitly presented. Then, in Section \ref{sec:box} we compute the induced box-shaped gamma-ray spectra and derive model-independent constraints on the narrow and wide box scenarios using Fermi-LAT \cite{fermilatsite} and H.E.S.S.~\cite{hesssite} data. Our final remarks are drawn in Section \ref{sec:conclusions}.

\section{Axion-mediated dark matter}
\label{sec:axion}

\par We consider the model presented in Refs.~\cite{axion1,axion15,axion2}, where the Standard Model (SM) is extended by a Dirac dark matter particle $\chi$ and a complex scalar $S$ charged under a global $U(1)_{\rm PQ}$ symmetry. Similar models were studied in Refs.~\cite{Nomura:2008ru,Mardon:2009gw} in the context of a dark matter explanation for the PAMELA positron excess. The Peccei-Quinn (PQ) transformations of the dark matter particle and the mediator are, respectively, $\chi\rightarrow e^{i\gamma_5 \alpha}\chi$ and $S\rightarrow e^{-2 i\alpha } S$. We further assume that $S$ acquires a vacuum expectation value such that the complex scalar field can be cast as $S=(v_s+s+ia)/\sqrt{2}$, where $v_s\equiv \sqrt{2} \,\langle S\rangle$ is the axion decay constant. The effective Lagrangian of the model then reads
\be
{\cal L}= i{\bar\chi}\gamma^\mu\partial_\mu\chi+|\partial_\mu S|^2-V(H,S)+{\cal L}_{\rm int}+{\cal L}_{\rm SM}\,,
\label{action}
\ee
where ${\cal L}_{\rm SM}$ is the Standard Model Lagrangian, which includes the Higgs potential $V_{\rm SM}=\lambda_H |H|^4+m^2_H |H|^2$, and 
\bea
V(H,S)&=& \lambda_S |S|^4+2\lambda_{HS}|S|^2 |H|^2+ m^2_S|S|^2-\Big(\frac{1}{2}m^{\prime 2}_S S^2+{\rm c.c.}\Big)\,, \label{potential} \\
{\cal L}_{\rm int}&=&-\lambda_\chi (S {\bar \chi} P_L\chi +S^* {\bar\chi}P_R \chi )+ \sum_{i=1,2}\frac{c_i\alpha_i}{ 8\pi  v_s}\, a F^i_{\mu\nu} {\tilde F}^{i\mu\nu}  \label{interaction}
\eea
with ${\tilde F}_{\mu\nu}\equiv\frac{1}{2}\epsilon_{\mu\nu\rho\sigma}F^{\rho\sigma}$. The constant parameters $c_i$ depend on the charges of the heavy fermions in the anomaly loop diagrams and are explicitly given in Refs.~\cite{axion15,axion2}. It is possible to show that the Lagrangian contains a mass matrix for the CP-even scalars which can be diagonalised by the field rotation \cite{axion1}
\be
s=\cos\tilde{\theta}\, {\tilde s}+\sin\tilde{\theta}\,{\tilde h}, \quad h=-\sin\tilde{\theta}\,{\tilde s}+\cos\tilde{\theta}\,{\tilde h} \, ,
\label{eq:eigenstates}
\ee
where the mixing angle is given by 
\be
\tan\,2\tilde{\theta}=\frac{2\lambda_{HS} v_s v}{\lambda_H v^2-\lambda_S v^2_s}\, ,
\label{eq:mixingangle}
\ee
$v_s$ and $v$ being the vacuum expectation values of the singlet and the Higgs doublet.

\par The axion has a mass $m_a=m^\prime_S$ which stems from the mass term  $\frac{1}{2}m^{\prime 2}_S S^2+ {\rm c.c.}$. This mass term can originate from the higher dimensional operator $\frac{g}{M^2_P} \,\Phi^4 S^2+{\rm c.c.}$, where $\Phi$ is a high-scale Peccei-Quinn breaking field, with $\langle \Phi\rangle=f_a\gg v_s$. The high-scale PQ breaking scale is constrained to be $10^9\,{\rm GeV}< f_a < 10^{12}\,{\rm GeV}$ from bounds of supernova cooling and relic density on the invisible axion coming from $\Phi$, respectively. Thus, the weak-scale axion mass $m_a$ ranges between $1\,{\rm GeV}$ and $10^6\,{\rm GeV}$, for $g\sim {\cal O}(1)$.

\par The interaction Lagrangian in Eq.~(\ref{interaction}) gives rise to a dark matter mass and to dark matter interactions with the CP-even scalar $s$ and the CP-odd scalar $a$ which, assuming for simplicity that the coupling $\lambda_\chi$ is real, are given by
\be
{\cal L}_{\rm DM}= -m_\chi \,{\bar \chi}\chi-\frac{1}{\sqrt{2}}\lambda_\chi s \,{\bar \chi}\chi-\frac{1}{\sqrt{2}}\lambda_\chi a\,{\bar\chi}\gamma^5 \chi\,,
\ee
where $m_\chi=\lambda_\chi v_s/\sqrt{2}$; the interaction term with the mass eigenstates $\tilde s,\tilde h$ can be easily obtained from Eq.~(\ref{eq:eigenstates}). 

\par Lastly, the effective axion interactions can be rewritten in terms of physical electroweak gauge bosons:
\be
{\cal L}_{\rm aVV}=\sum_{i\leq j}c_{V_i V_j}a\,\epsilon_{\mu\nu\rho\sigma} F^{\mu\nu}_{V_i}F^{\rho\sigma}_{V_j}\,, \label{anomalyint}
\ee
where the only non-vanishing terms are
\bea
c_{\gamma\gamma}&=& \frac{1}{16\pi  v_s}(c_1\alpha_1 \cos^2\theta_W+c_2\alpha_2 \sin^2\theta_W)\,, \nonumber \\
c_{Z\gamma}&=& \frac{1}{16\pi v_s} (c_2\alpha_2 -c_1\alpha_1)\sin(2\theta_W)\,, \nonumber \\
c_{ZZ}&=& \frac{1}{16\pi v_s}(c_2\alpha_2  \cos^2\theta_W+c_1 \alpha_1 \sin^2\theta_W)\,,\nonumber \\
c_{WW}&=& \frac{ c_2\alpha_2}{8\pi v_s}\,.
\label{eq:decay-constants}
\eea 

\par With these interactions, the dark matter particle is absolutely stable due to the residual $Z_2$ symmetry which emerges after the Peccei-Quinn symmetry is spontaneously broken by the singlet vacuum expectation value. However, the CP-even singlet scalars $\tilde{s},\tilde{h}$ can decay into SM particles due to the mixing with the Higgs boson. Additionally, the heavier eigenstate $\tilde{s}$ may decay into a pair of the lighter eigenstates $\tilde{h}$, and both eigenstates $\tilde{s},\tilde{h}$ may decay into a pair of dark matter particles if kinematically allowed. 
The decay widths for all these processes can be found in Ref.~\cite{axion1}.
We also note that the scalar eigenstates may also decay into a pair of axions if kinematically allowed.

\par The axion, on the other hand, can decay into two electroweak gauge bosons with rates given by
\bea
\Gamma_a(\gamma\gamma)&=&\frac{m^3_a}{\pi} |c_{\gamma\gamma}|^2\,, \nonumber \\
\Gamma_a(Z\gamma)&=& \frac{m^3_a}{2\pi}|c_{Z \gamma }|^2\Big(1-\frac{m^2_Z}{m^2_a}\Big)^3\,, \nonumber\\
\Gamma_a(ZZ)&=& \frac{m^3_a}{\pi} |c_{ZZ}|^2\Big(1-\frac{4m^2_Z}{m^2_a}\Big)^{3/2}\,, \nonumber\\
\Gamma_a(W^+W^-)&=&  \frac{m^3_a}{2\pi}|c_{WW }|^2\Big(1-\frac{4m^2_W}{m^2_a}\Big)^{3/2}\,.
\label{eq:decay-rates}
\eea
Decays into Standard Model fermions are not allowed at tree level.

\begin{figure}[t]
\centering%
\includegraphics[width=0.32\textwidth]{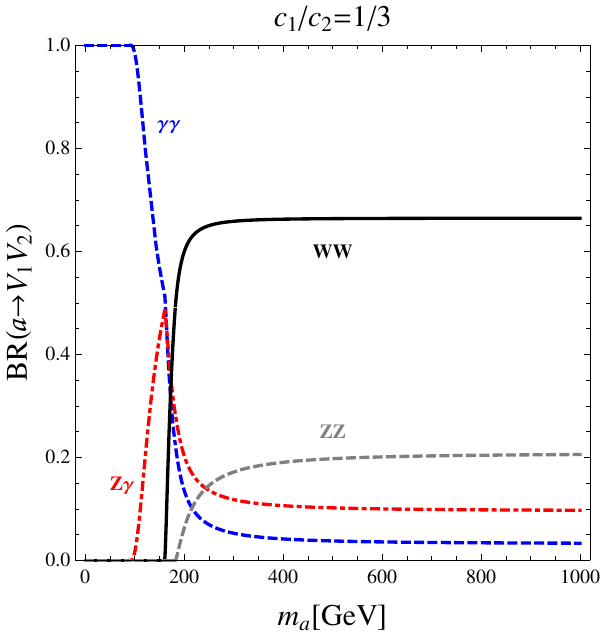} 
\includegraphics[width=0.32\textwidth]{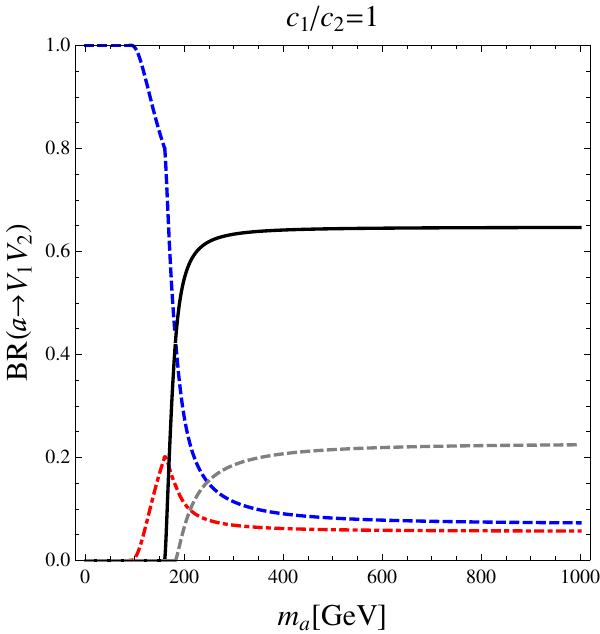} 
\includegraphics[width=0.32\textwidth]{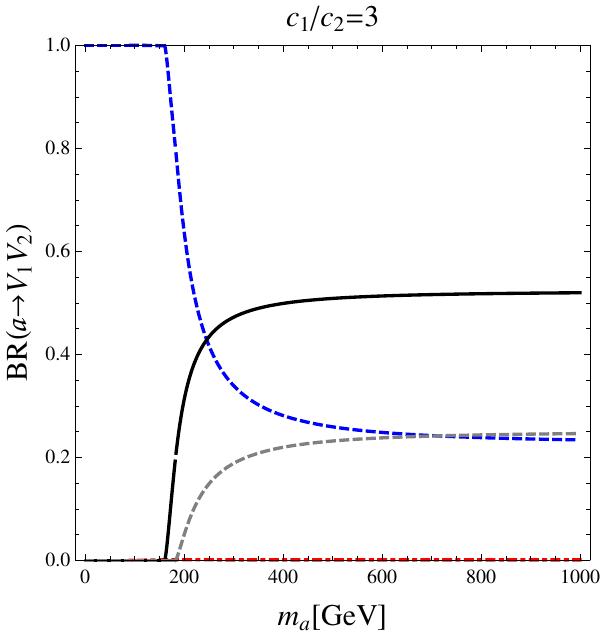}
\caption{\small Branching ratios of the axion decay as a function of the axion mass $m_a$ for a ratio of anomaly coefficients $c_1/c_2=1/3, 1, 3$, when the axion couples to vector-like doublet ($Y=\pm1/2$) $+$ triplet ($Y=0$), vector-like doublet ($Y=\pm1/2$), and vector-like doublet ($Y=\pm1/2$) $+$ vector-like singlet ($Y=\pm1$) fermions, respectively. Solid black, dashed grey, dashed blue and dot-dashed red lines correspond to the $WW, ZZ, \gamma\gamma, Z\gamma$ decay modes, respectively.}\label{fig:BRvsma}
\end{figure}

\par The branching ratios for the axion decay as a function of the axion mass are shown in Fig.~\ref{fig:BRvsma} for various choices of $c_1/c_2$. It is apparent from the figure and from Eqs.~(\ref{eq:decay-constants},\ref{eq:decay-rates}) that, for generic choices of parameters, all decay channels have sizable branching ratios provided they are kinematically open. More specifically, the decay channels  $a\rightarrow \gamma\gamma$ and $a\rightarrow Z\gamma$ can have large branching ratios and even be the dominant decay channels. This property of the model will have important implications for indirect searches, as will be discussed in detail in Section \ref{sec:box}.

\section{Tree-level annihilation cross sections}
\label{sec:annihilations}

\par At tree level there are three possible dark matter t-channel annihilations, $\chi{\bar\chi}\rightarrow as, aa, ss$, which easily dominate over the one loop induced s-channel annihilations into gauge bosons $\chi \bar\chi \rightarrow V_1 V_2$ unless $m_a\sim 2 m_\chi$, namely when the rate for the latter processes is resonantly enhanced. In this section we will derive the velocity-averaged annihilation cross sections for these processes assuming for simplicity that the mixing between the two CP-even scalars is zero; the full expressions including the Higgs-singlet mixing can be easily derived from the formulas below using the mixing angle defined in Eq.~(\ref{eq:mixingangle}) and the mass eigenvalues corresponding to $\tilde{s}$ and $\tilde{h}$.

\par The dark matter annihilation cross sections times relative velocity (without thermal average) for all the t-channel processes are given by
\bea \label{eq:fullsigmav}
\left(\sigma v\right)_{ij}=
\frac{1}{32 \pi ^2\kappa s}
 \left(1-\frac{\left(m_i-m_j\right){}^2}{s}\right)^{1/2}
 \left(1-\frac{\left(m_i+m_j\right){}^2}{s}\right)^{1/2}
 \int\D\Omega \overline{|\mathcal{M}|^2_{ij}}
\eea
with $\kappa=1$ or $2$ for the annihilation channels $ij=as$ or $ij=aa, ss$, respectively. For later convenience, we shall use the Mandelstam variables $s, t, u$: 
\begin{eqnarray}
 s&=&4m_\chi^2\left( 1-\frac{v_\text{rel}^2}{4}\right)^{-1}\simeq 4m_\chi^2 \left(1+\frac{v_\text{rel}^2}{4} \right)\,,\\ \nonumber
 t&=& \frac{1}{2}\left(m_i^2 + m_j^2 +2m_\chi^2 -s\right) + \frac{v_{\text{rel}}}{4}\cos{\theta} \Big[ \left(m_i^2 - m_j^2 +s \right)^2 -4s m_i^2\Big]^{1/2}  \\
 &\simeq& \frac{1}{2}
   \left(m_i^2+m_j^2-2 m_\chi^2-m_\chi^2 v_{\text{rel}}^2 \right)+
   \frac{v_{\text{rel}}}{4}\cos{\theta}  \Big[ \left(m_i^2-m_j^2+4m_\chi^2\right)^2-16m_\chi^2 m_i^2 \Big]^{1/2} \,,\\ \nonumber
 u&=& \frac{1}{2}\left(m_i^2 + m_j^2 +2m_\chi^2-s \right) - \frac{v_{\text{rel}}}{4}\cos{\theta} \Big[ \left(m_i^2 - m_j^2 +s\right)^2 -4s m_i^2\Big]^{1/2} \\
 &\simeq& \frac{1}{2}
   \left(m_i^2+m_j^2-2 m_\chi^2-m_\chi^2 v_{\text{rel}}^2\right)-
   \frac{v_{\text{rel}}}{4}\cos{\theta} \Big[ \left(m_i^2-m_j^2+4m_\chi^2\right)^2-16m_\chi^2 m_i^2 \Big]^{1/2}  \,,
\end{eqnarray}
where $v_{\text{rel}}$ is the M\"oller velocity, $\theta$ is the angle between the direction of the incoming dark matter particle and the outgoing axion or scalar, both expressed in the centre of mass frame, and the approximate expressions are valid in the non-relativistic limit up to order $\mathcal{O}(v_\text{rel}^2)$. Besides, the amplitudes squared are
\begin{eqnarray}
\label{eq:ampl1}
\overline{|\mathcal{M}|^2_{as}}&=&\frac{|\lambda_\chi|^4}{8}
\frac{(t+u-2m_\chi^2)^2}{(t-m_\chi^2)^2(u-m_\chi^2)^2} \Big[\left(m_s^2+m_\chi^2-t\right) \left(m_s^2+m_\chi^2 -u \right)- m_s^2\,s+4 m_a^2 m_\chi^2 \Big]\,, \\ \label{eq:ampl2}
\overline{|\mathcal{M}|^2_{aa}}&=&\frac{|\lambda_\chi|^4}{8}
\left(\frac{1}{t-m_\chi^2}-\frac{1}{u-m_\chi^2}\right)^2
   \Big[\left(m_a^2+m_\chi^2 -t \right) \left(m_a^2+m_\chi^2 -u\right)-m_a^2\,s\Big]\,,  \\ \nonumber
\overline{|\mathcal{M}|^2_{ss}}&=&\frac{|\lambda_\chi|^4}{8}
\frac{1}{(t-m_\chi^2)^2 (u-m_\chi^2)^2}\Big[(u-t)^2 \Big(\left(m_s^2+m_\chi^2-t\right)
   \left(m_s^2+m_\chi^2-u \right)-m_s^2\,s \\ \label{eq:ampl3}
   && +4m_\chi^2(t+u-2m_\chi^2)\Big) +4m_\chi^2(s-4m_\chi^2) (t+u-2m_\chi^2)^2\Big]\,.
\end{eqnarray}

\par The annihilation cross sections times velocity for the three annihilation channels in the non-relativistic regime read
\begin{eqnarray}
 \left(\sigma v_\text{rel}\right)_{as}&\simeq & 
 \frac{|\lambda_\chi|^4}{64\pi m_\chi^2}
 \frac{\left(m_a^2-m_s^2+4m_\chi^2\right)^2}{\left(m_a^2+m_s^2-4m_\chi^2 \right)^2}
 \left( 1-\frac{\left(m_a-m_s\right)^2}{4m_\chi^2}\right)^{1/2}
 \left(1-\frac{\left(m_a+m_s\right)^2}{4m_\chi^2}\right)^{1/2} \,, \\
 \left(\sigma v_\text{rel}\right)_{aa}&\simeq &
 \frac{|\lambda_\chi|^4}{96\pi}
 \frac{m_\chi^6}{\left(m_a^2-2 m_\chi^2\right)^4}
 \left(1-\frac{m_a^2}{m_\chi^2}\right)^{5/2}v_\text{rel}^2 \,, \\
 \left(\sigma v_\text{rel}\right)_{ss}&\simeq & 
 \frac{|\lambda_\chi|^4}{96\pi}
 \frac{m_\chi^2}{\left(m_s^2-2m_\chi^2 \right)^4}
 \left(2\left(m_s^2-2m_\chi^2 \right)^2 + m_\chi^4 \right)
 \left(1-\frac{m_s^2}{m_\chi^2}\right)^{1/2} 
 v_\text{rel}^2 \,.
\end{eqnarray}
Clearly, the process $\chi\bar{\chi}\to as$ is $s$-wave, whereas the processes $\chi\bar{\chi}\to aa,\, ss$ are both $p$-wave. Therefore, the annihilation today ($v_{\textrm{rel}}\sim 10^{-3}$) is dominated by the $as$ channel if this is kinematically accessible and, if not, by either the $aa$ or the $ss$ channels. Notice that, if the $as$ channel is kinematically forbidden, either only the $aa$ channel or only the $ss$ channel are accessible. To summarise, ${\rm BR}(\chi\bar{\chi}\to as,aa,ss)\simeq 1,0,0$ for $m_a+m_s < 2m_\chi$; ${\rm BR}(\chi\bar{\chi}\to as,aa,ss)\simeq 0,1,0$ for $m_a<m_\chi$ and $m_a+m_s > 2m_\chi$; and ${\rm BR}(\chi\bar{\chi}\to as,aa,ss)\simeq 0,0,1$ for $m_s<m_\chi$ and $m_a+m_s > 2m_\chi$. The thermally averaged cross sections are obtained from the full cross section expressions Eqs.~\eqref{eq:fullsigmav} and (\ref{eq:ampl1}--\ref{eq:ampl3}) by applying the full averaging procedure (see, for instance, Eq.~(3.8) in Ref.~\cite{Gondolo:1990dk}) with a freeze-out temperature $T_{\textrm{fo}}=m_\chi/25$. 

\par We include for completeness the s-channel dark matter annihilation cross sections into electroweak gauge bosons. The corresponding cross sections with axion mediator for a photon pair and $Z\gamma$ are given by \cite{axion1,axion2}
\bea
\langle\sigma v\rangle_{\gamma\gamma}&=&\frac{ 1}{2\pi}|\lambda_\chi|^2 |c_{\gamma\gamma}|^2\,  
\frac{16 m^4_\chi}{(4m^2_\chi-m^2_a)^2+\Gamma^2_a m^2_a}\,,\label{axionxsection} \\
\langle\sigma v\rangle_{Z\gamma}&=&\frac{1}{4\pi}|\lambda_\chi|^2 |c_{Z \gamma }|^2\,   \,
\frac{16 m^4_\chi}{(4m^2_\chi-m^2_a)^2+\Gamma^2_a m^2_a}\Big(1-\frac{m^2_Z}{4 m^2_\chi}\Big)^3\, ,
\eea
which can be sizable when $m_a\simeq 2m_\chi$. In the above expressions $\Gamma_a$ is the axion total decay width. The cross sections for the other s-channels with axion mediator are\footnote{The typos due to a factor 8 in the annihilation cross sections in Refs.~\cite{axion1,axion15} were corrected in Ref.~\cite{axion2}.}
 \bea
 \langle \sigma v\rangle_{ZZ}&=&  \frac{1}{2\pi} |\lambda_\chi|^2|c_{Z Z}|^2\,\frac{16m^4_\chi}{(4m^2_\chi-m^2_a)^2+m^2_a\Gamma^2_a}\Big(1-\frac{m^2_Z}{m^2_\chi}\Big)^{3/2}\,,\\
  \langle \sigma v\rangle_{W W}&=&\frac{1}{4\pi} |\lambda_\chi|^2|c_{WW}|^2\,\frac{16m^4_\chi}{(4m^2_\chi-m^2_a)^2+m^2_a\Gamma^2_a}\Big(1-\frac{m^2_W}{m^2_\chi}\Big)^{3/2}\,.
\eea
There are in addition s-channel annihilation channels mediated by CP-even scalars. However, as argued in Refs.~\cite{axion1,axion15}, these are p-wave suppressed and thus irrelevant for indirect dark matter detection. These channels could though be important in determining the relic density at freeze-out when the CP-even scalar singlet has a sizable mixing with the Standard Model Higgs \cite{axion1,axion15}.

\par In this paper we will be interested in the regime where $m_a < 2 m_\chi$ and will assume a small mixing between the CP-even scalar and the Standard Model Higgs. Therefore, the relic density is essentially determined by the annihilation cross sections of the processes $\chi \bar \chi\rightarrow as,aa,ss$. As will be shown in the next section, depending on the dark matter coupling to the axion $\lambda_\chi$ as well as on $m_\chi$, $m_a$ and $m_s$, it is possible to reproduce the measured relic density for a wide range of masses for reasonable values of the coupling $\lambda_\chi$ and without any fine-tuning of parameters.

\section{Box-shaped gamma-ray signatures}
\label{sec:box}

\par The annihilation channels $\chi{\bar\chi}\rightarrow as$ and $\chi{\bar\chi}\rightarrow aa$ produce prominent gamma-ray features from the decay in flight of the axion into two photons or into one photon and one $Z$ boson. Let us discuss first the energy spectrum produced in the annihilation $\chi{\bar\chi}\rightarrow as$. The decay of the axion $a\rightarrow \gamma \gamma$ produces two photons with identical energy in the rest frame of the axion, $E^{(\gamma\gamma)}_{{\gamma,\rm  RF}}=m_a/2$. However, in the Galactic frame, where the dark matter particles move non-relativistically, the photon energy reads
\bea
E_\gamma^{(\gamma\gamma)}=\frac{1}{\gamma} E^{(\gamma\gamma)}_{{\gamma,\rm  RF}} (1-v_a \cos\theta )^{-1}\,,
\eea
where $\gamma\equiv 1/\sqrt{1-v^2_a}$, $v_a$ is the axion velocity in $c$ units and $\theta$ is the angle between the direction of the axion and the direction of the photon. Concretely, for the $\chi{\bar\chi}\rightarrow as$ channel the velocity of the axion is
\be
v_a=\frac{p_a}{E_a}=\sqrt{1-\frac{m^2_a}{m^2_\chi}\Big(1 +\frac{m^2_a-m^2_s}{4m_\chi^2}\Big)^{-2}}\,,
\ee
which gives a photon energy in the Galactic frame
\be
E_\gamma^{(\gamma\gamma)}=\frac{m^2_a}{2m_\chi}\Big(1 +\frac{m^2_a-m^2_s}{4m_\chi^2}\Big)^{-1}\left(1-\cos\theta \sqrt{1-\frac{m^2_a}{m^2_\chi}\bigg(1 +\frac{m^2_a-m^2_s}{4m_\chi^2}\bigg)^{-2}} \right)^{-1}\,.
\label{eq:energy-gg}
\ee
Since the axion decays isotropically, the resulting energy spectrum presents a box-shape structure with the photon energy ranging from $E^{(\gamma\gamma)}_-$ to $E^{(\gamma\gamma)}_+$, where $E^{(\gamma\gamma)}_\pm=\frac{1}{2}A m_\chi(1\pm \sqrt{1-\frac{m^2_a}{A^2m^2_\chi} })$ and $A=1+(m^2_a-m^2_s)/(4 m^2_\chi)$. Analogously, the energy of the photon produced in the annihilation $\chi{\bar \chi}\rightarrow a s$ followed by the decay  $a\rightarrow Z\gamma$  reads
\be
\label{eq:energy-gZ}
E_\gamma^{(Z\gamma)}=\frac{m^2_a}{2m_\chi}\Big(1-\frac{m^2_Z}{m^2_a}\Big)\Big(1 +\frac{m^2_a-m^2_s}{4m_\chi^2}\Big)^{-1}\left(1-\cos\theta \sqrt{1-\frac{m^2_a}{m^2_\chi}\bigg(1 +\frac{m^2_a-m^2_s}{4m_\chi^2}\bigg)^{-2}} \right)^{-1}\,,
\ee
producing also a box-shaped spectrum with photon energies ranging from $E^{(Z\gamma)}_-$ to $E^{(Z\gamma)}_+$, where $E^{(Z\gamma)}_\pm=\frac{1}{2}A m_\chi \Big(1-\frac{m^2_Z}{m^2_a}\Big) \Big(1\pm \sqrt{1-\frac{m^2_a}{A^2m^2_\chi} }\Big)$. Therefore, the energy spectrum of hard photons produced in the annihilation channel $\chi {\bar \chi}\rightarrow a s$ is
\begin{eqnarray}
\frac{dN_\gamma}{dE_\gamma}&=&\frac{2}{E^{(\gamma\gamma)}_+-E^{(\gamma\gamma)}_-}\Theta(E_\gamma-E^{(\gamma\gamma)}_-)\Theta(E^{(\gamma\gamma)}_+-E_\gamma) {\rm BR}(a\rightarrow \gamma\gamma)\nonumber \\
&&\quad +\frac{1}{E^{(Z\gamma)}_+-E^{(Z\gamma)}_-}\Theta(E_\gamma-E^{(Z\gamma)}_-)\Theta(E^{(Z\gamma)}_+-E_\gamma) {\rm BR}(a\rightarrow Z\gamma)\,,
\label{eq:spectrum-as}
\end{eqnarray}
where $\Theta$ is the Heaviside function. In the presence of mixing between the singlet scalar and the Higgs boson, there are two annihilation channels: $\chi{\bar\chi}\rightarrow a \tilde s$ and $\chi{\bar\chi}\rightarrow a \tilde h$. In that case, the spectrum can be derived from the previous expressions substituting  $m_s$ by the mass eigenvalues associated to $\tilde{s}$ and $\tilde{h}$. For a small Higgs mixing, the dominant annihilation channel is $\chi{\bar\chi}\rightarrow a \tilde s$. 
In our analysis we assume that this is the case.

\par When the dark matter particle annihilates in the channel $\chi{\bar\chi}\rightarrow aa$, the energy of the photon produced in the decay in flight of the axion can be straightforwardly calculated from Eqs.~(\ref{eq:energy-gg},\ref{eq:energy-gZ}) upon replacing  $m_s$ by $m_a$, namely
\be
E_\gamma^{(\gamma \gamma)}=\frac{m^2_a}{2m_\chi}\left(1-\cos\theta \sqrt{1-\frac{m^2_a}{m^2_\chi}} \right)^{-1}
\ee
for the case of the decays $a\rightarrow \gamma \gamma$ and
\be
E_\gamma^{(Z\gamma)}=\frac{m^2_a}{2m_\chi}\Big(1-\frac{m^2_Z}{m^2_a}\Big)\left(1-\cos\theta \sqrt{1-\frac{m^2_a}{m^2_\chi}} \right)^{-1}
\ee
for the case of the decays $a\rightarrow Z\gamma$. The energy spectrum of photons produced in the annihilations is then given by Eq.~(\ref{eq:spectrum-as}) with an additional factor of two due to the fact that two axions are produced in each annihilation.

\par Finally, the annihilations of Dirac dark matter particles produce a gamma-ray flux \cite{Bertone:2004pz,Bergstrom:2012fi}
\begin{equation}\label{unconvol}
\frac{d \Phi_\gamma}{dE_\gamma}= \frac{1}{16\pi m_\chi^2} \, \sum_f \langle \sigma v \rangle_f \frac{dN^f_{\gamma}}{dE_\gamma} \, \frac{1}{\Delta \Omega} \int_{\Delta \Omega}{d\Omega \, J_{\rm ann}} \quad ,
\end{equation}
where $\langle \sigma v \rangle_f$ is the velocity-averaged annihilation cross section in the channel $f$, $\Delta \Omega$ is the observed field of view and $J_{\rm ann}=\int_{l.o.s.}{ds \, \rho_{\rm DM}^2}$ is the integral of the squared dark matter density $\rho_{\rm DM}$ along the line of sight. The convolution of the flux in Eq.~(\ref{unconvol}) with the experimental energy resolution is done following Ref.~\cite{Ibarra:2012dw} (see details below).

\par
\begin{figure}[t]
\centering%
\includegraphics[width=7.5cm]{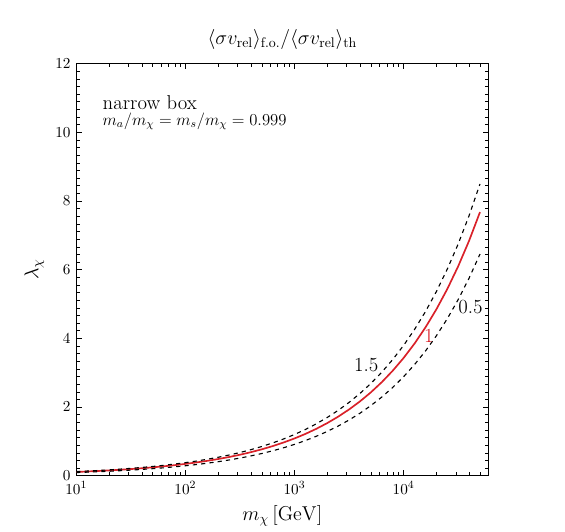} 
\includegraphics[width=7.5cm]{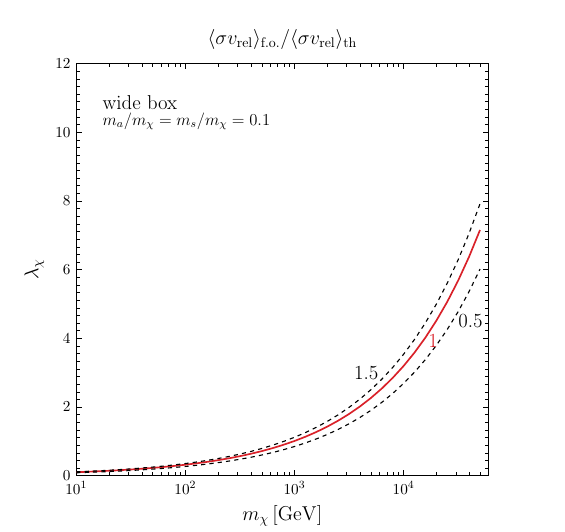}
\caption{\small Contours of the total annihilation cross section at freeze-out $\langle\sigma v\rangle_{\rm fo}$ in units of the thermal cross section for Dirac dark matter particles, $\langle\sigma v\rangle_{\rm th}=6\times 10^{-26}\,{\rm cm}^3 \,{\rm s}^{-1}$, as a function of $m_\chi$ and $\lambda_\chi$. The left (right) panel shows the narrow (wide) box scenario with $m_a/m_\chi=m_s/m_\chi=0.999$ ($0.1$).
}
\label{fig:lchi-vs-mchi}
\end{figure}

\begin{figure}[h]
\centering%
\includegraphics[width=7.5cm]{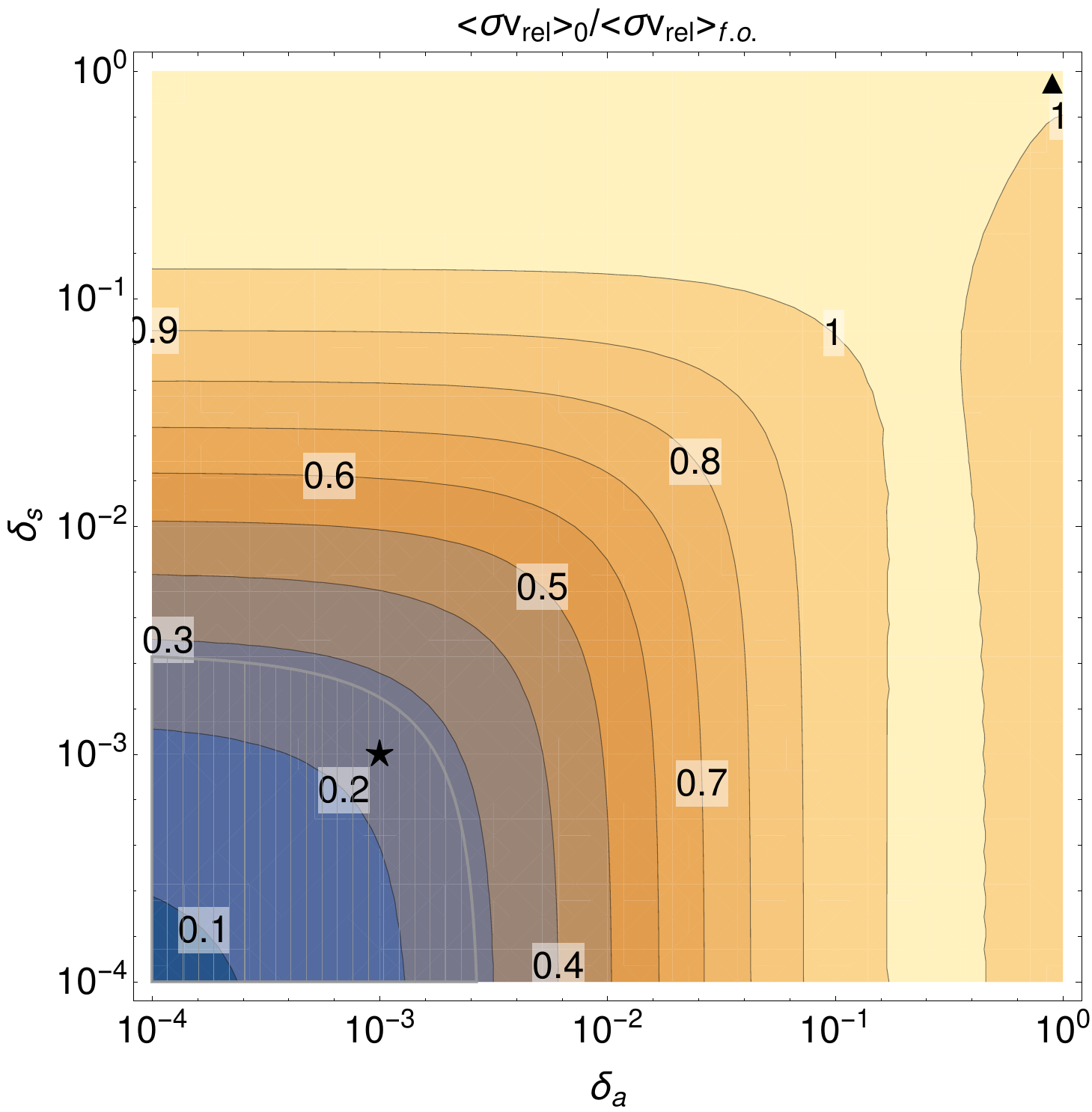}
\caption{\small Contours of the ratio of the present to thermal total annihilation cross sections $\langle\sigma v\rangle_0/\langle\sigma v\rangle_{\rm fo}$ as a function of $\delta_a\equiv (m_\chi-m_a)/m_\chi$ and $\delta_s\equiv (m_\chi-m_s)/m_\chi$. The hatched region marks the parameter space where the width of the box is $10\%$ or less than the maximum energy of the photons. We also show the location of our exemplary points with $m_a/m_\chi=m_s/m_\chi=0.999$ for the narrow box scenario ($\bigstar$) and $0.1$ for the wide box scenario ($\blacktriangle$).
}
\label{fig:thermalrelic}
\end{figure}

 For a given annihilation channel, the energy spectrum of photons depends on $m_a/m_\chi$, $m_s/m_\chi$, ${\rm BR}(a\rightarrow \gamma\gamma)$ and ${\rm BR}(a\rightarrow Z\gamma)$. As discussed in the previous section, the cross sections for the annihilations $\chi\bar \chi\rightarrow as, aa, ss$ relative to the total cross section are ${\rm BR}(\chi\bar{\chi}\to as,aa,ss)\simeq 1,0,0$ for $m_a+m_s < 2m_\chi$; ${\rm BR}(\chi\bar{\chi}\to as,aa,ss)\simeq 0,1,0$ for $m_a<m_\chi$ and $m_a+m_s > 2m_\chi$; and ${\rm BR}(\chi\bar{\chi}\to as,aa,ss)\simeq 0,0,1$ for $m_s<m_\chi$ and $m_a+m_s > 2m_\chi$. One can then identify two possible scenarios depending on the width of the box compared to the experimental energy resolution: the ``narrow box'' scenario and the ``wide box'' scenario.  In the former case, the axion is produced almost at rest and the gamma-ray spectrum resembles a line, while in the latter the axion is very relativistic and the photons produced in its decay present a large spread in energy. In both cases the measured relic density can be reproduced for a wide range of dark matter masses with $\lambda_\chi\sim{\cal O}(1)$, as demonstrated in Fig.~\ref{fig:lchi-vs-mchi}. In Fig.~\ref{fig:thermalrelic}, we also depict the ratio of the present to thermal total annihilation cross sections, $\langle\sigma v\rangle_0/\langle\sigma v\rangle_{\rm fo}$. We find that the narrower the gamma-ray box becomes, the more sizeable the velocity-dependent contribution to the dark matter annihilation cross section and the smaller the ratio of the present to thermal cross sections are. For instance, we get $\langle\sigma v\rangle_0/\langle\sigma v\rangle_{\rm fo}\simeq 0.238$ for a narrow box with $m_a/m_\chi=m_s/m_\chi=0.999$, while $\langle\sigma v\rangle_0/\langle\sigma v\rangle_{\rm fo}\simeq 1.013$ for a wide box with $m_a/m_\chi=m_s/m_\chi=0.1$.

\par
\begin{figure}[t]
\centering%
\includegraphics[width=7.5cm]{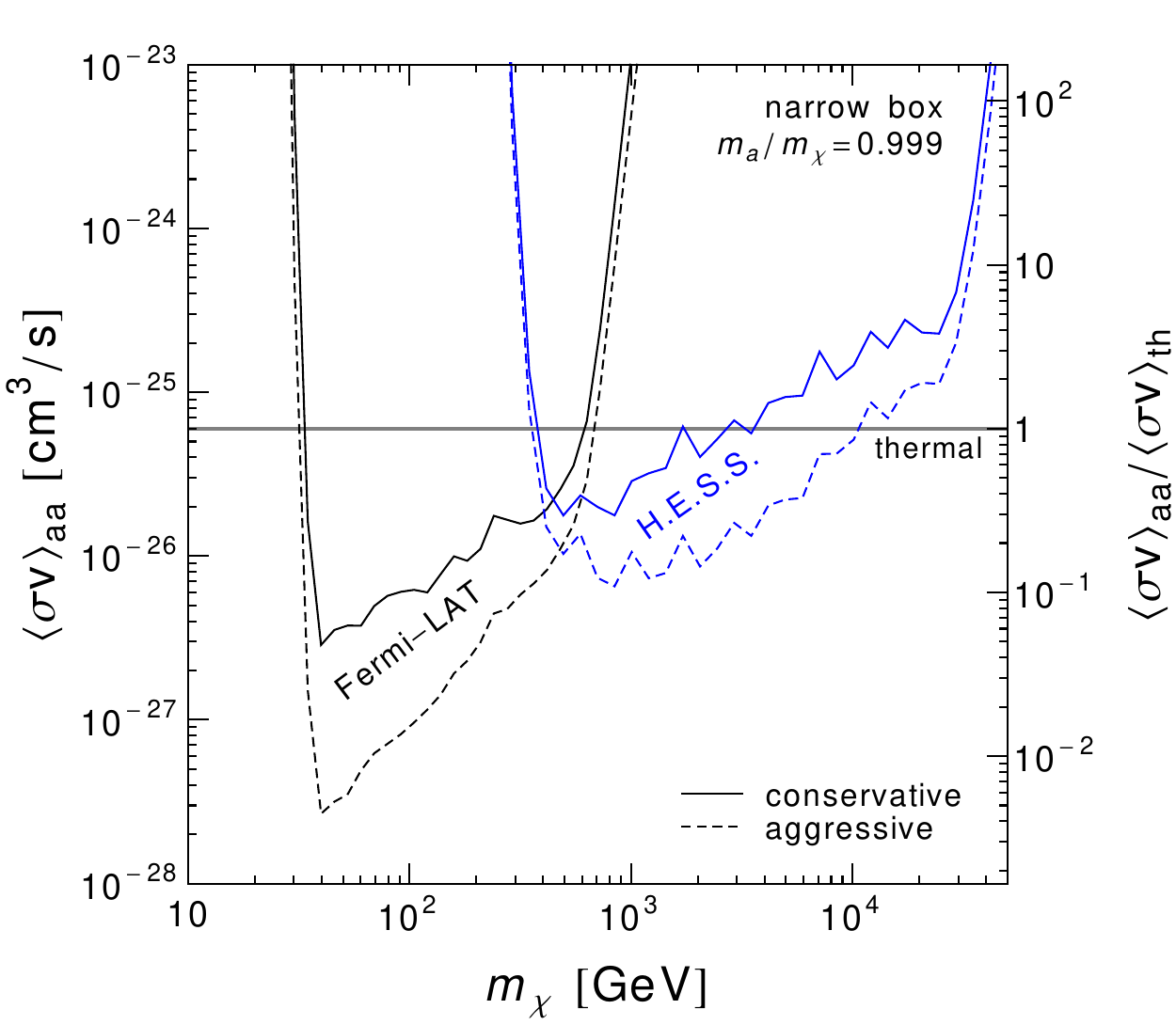} 
\includegraphics[width=7.5cm]{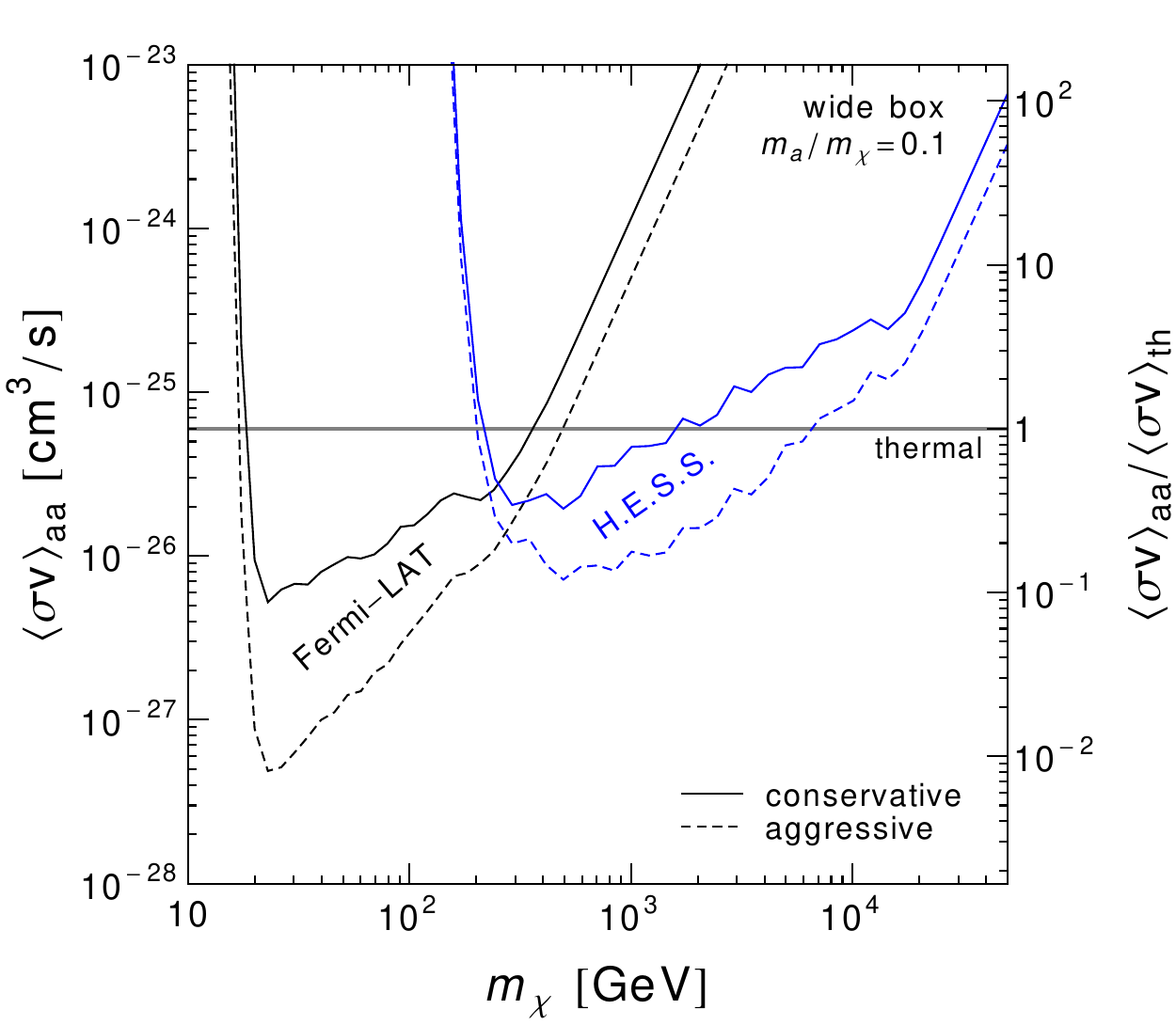}
\caption{\small Upper limits on the annihilation cross section in the channel $\chi \bar \chi\rightarrow a a$ from Fermi-LAT observations of the Galactic centre region \cite{Weniger:2012tx} and from H.E.S.S. observations of the Galactic ridge region \cite{Aharonian:2006au}, assuming ${\rm BR}(a\rightarrow \gamma\gamma)=1$. The left (right) panel shows the narrow (wide) box scenario with $m_a/m_\chi=m_s/m_\chi=0.999$ ($0.1$). The solid and dashed curves indicate, respectively, the limits following the conservative and aggressive approaches described in the text. The thermal cross section for Dirac dark matter particles, $\langle\sigma v\rangle_{\rm th}=6\times 10^{-26}\,{\rm cm}^3 \,{\rm s}^{-1}$, is shown by the horizontal line.
}
\label{fig:limits}
\end{figure}

 We have calculated limits on the annihilation cross section as a function of the dark matter mass for the process $\chi \bar \chi \rightarrow a a$ in the narrow and wide box scenarios from the Fermi-LAT observations of the Galactic centre region (Reg3 in \cite{Weniger:2012tx}) and from the H.E.S.S. observations of the Galactic ridge region \cite{Aharonian:2006au}. The energy resolution of Fermi-LAT is taken from Refs.~\cite{Rando:2009yq,fermilatsite2}, while for H.E.S.S.~a constant relative energy resolution of 15\% is assumed. Besides, for the dark matter distribution we adopt the Einasto profile, favoured by recent $N$-body simulations \cite{Navarro:2008kc,Hayashi:2007uk,Gao:2007gh},
\be
\rho_{\rm DM}(r)\propto \exp\left[-\frac{2}{\alpha}\left(\frac{r}{r_s}\right)^\alpha\right]
\ee
with  $r_s=20$ kpc and $\alpha=0.17$ \cite{Navarro:2003ew,Springel:2008cc} and normalised to $\rho_{\rm DM}(r=8.5\textrm{ kpc})=0.4\,{\rm GeV}\,{\rm cm}^{-3}$ \cite{Catena:2009mf,Weber:2009pt,Salucci:2010qr,Pato:2010yq}. The J-factors read $\frac{1}{\Delta\Omega}\int_{\Delta \Omega}{d\Omega \, J_{\rm ann}}\simeq 3.02\times 10^{23}\,{\rm GeV}^2\,{\rm cm}^{-5}$ for the Galactic centre region (Reg3 in \cite{Weniger:2012tx}) and $\frac{1}{\Delta\Omega}\int_{\Delta \Omega}{d\Omega \, J_{\rm ann}}\simeq 1.29\times 10^{25}\,{\rm GeV}^2\,{\rm cm}^{-5}$ for the Galactic ridge region defined in \cite{Aharonian:2006au}. The 2$\sigma$ limits are shown in Fig.~\ref{fig:limits}, assuming for concreteness $m_a/m_\chi=m_s/m_\chi=0.999$ in the case of the narrow box scenario and $m_a/m_\chi=m_s/m_\chi=0.1$ for the wide box scenario (choosing a different mass for $a$ and $s$ would lead to very similar phenomenology). These values for the mass ratios are indicated in Fig.~\ref{fig:thermalrelic} with a star and a triangle, respectively. Furthermore, in Fig.~\ref{fig:limits} we assume ${\rm BR}(a\rightarrow\gamma\gamma)=1$, which holds to a good approximation when $m_a<m_W$, as commonly occurs in the wide box scenario, but the limits can be trivially rescaled for a different branching ratio. We have computed the limits following two different approaches to bracket the uncertainty regarding background modelling: the solid curves labelled ``conservative'' correspond to the limits assuming that the background flux is zero, while the dashed curves labelled ``aggressive'' assume that the background meets the central points of the measurement. Let us note that these limits apply not only to the model discussed in this paper but to any model where the dark matter particle is a Dirac fermion that annihilates into two (pseudo-)scalars, which in turn decay in flight into two photons. In the case of a Majorana dark matter particle, the limits in Fig.~\ref{fig:limits} are stronger by a factor of two (and the thermal cross section is reduced by a factor of two). The limits corresponding to the decay of the (pseudo-)scalars into $Z\gamma$ can be readily derived from these plots with the appropriate substitutions in the kinematics, as well as the limits for the annihilation channel $\chi \bar{\chi}\rightarrow a s$. Note that Fig.~\ref{fig:limits} shows the limits for the annihilation channel $\chi{\bar\chi}\rightarrow aa$, which is not realised in the narrow and wide boxes we consider here. For those benchmark models, where the only relevant channel is now $\chi{\bar\chi}\rightarrow as$, the model-independent limits are a factor 2 weaker than the ones plotted in the Fig.~\ref{fig:limits} (assuming $m_s=m_a$).

\par We illustrate the potential of axion-mediated dark matter to produce observable signatures by showing the predicted gamma-ray flux for two exemplary points, one for the narrow box scenario and one for the wide box scenario, which produce an excess at around $130$ GeV, motivated by the recent hint of a sharp feature at this energy \cite{Bringmann:2012vr,Weniger:2012tx}. Although it is tempting to try to explain this signal in the framework of our model, we will not attempt to make a fit to the data.

\par
\begin{figure}[t]
\centering%
\includegraphics[width=7.5cm]{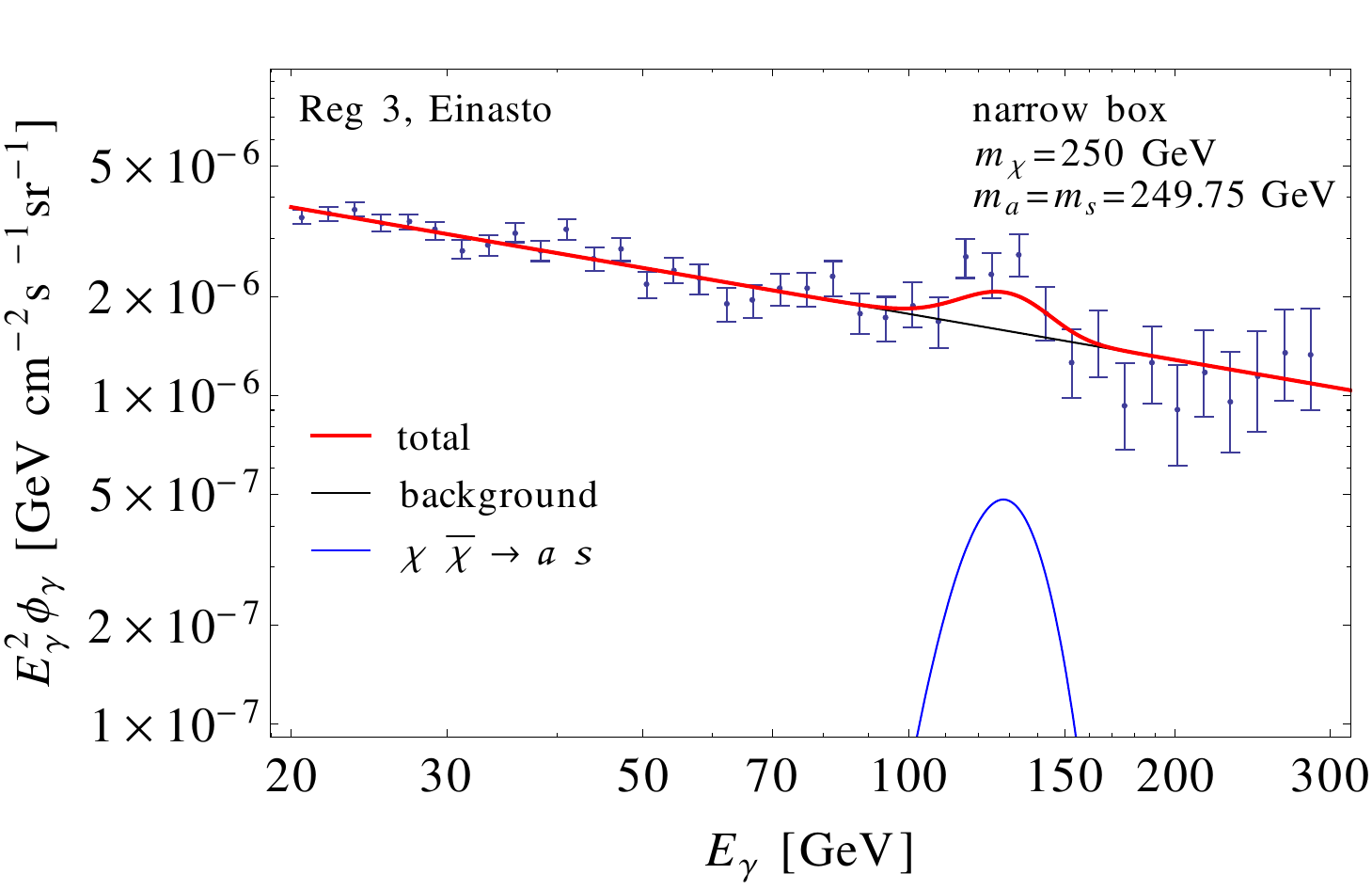} 
\includegraphics[width=7.5cm]{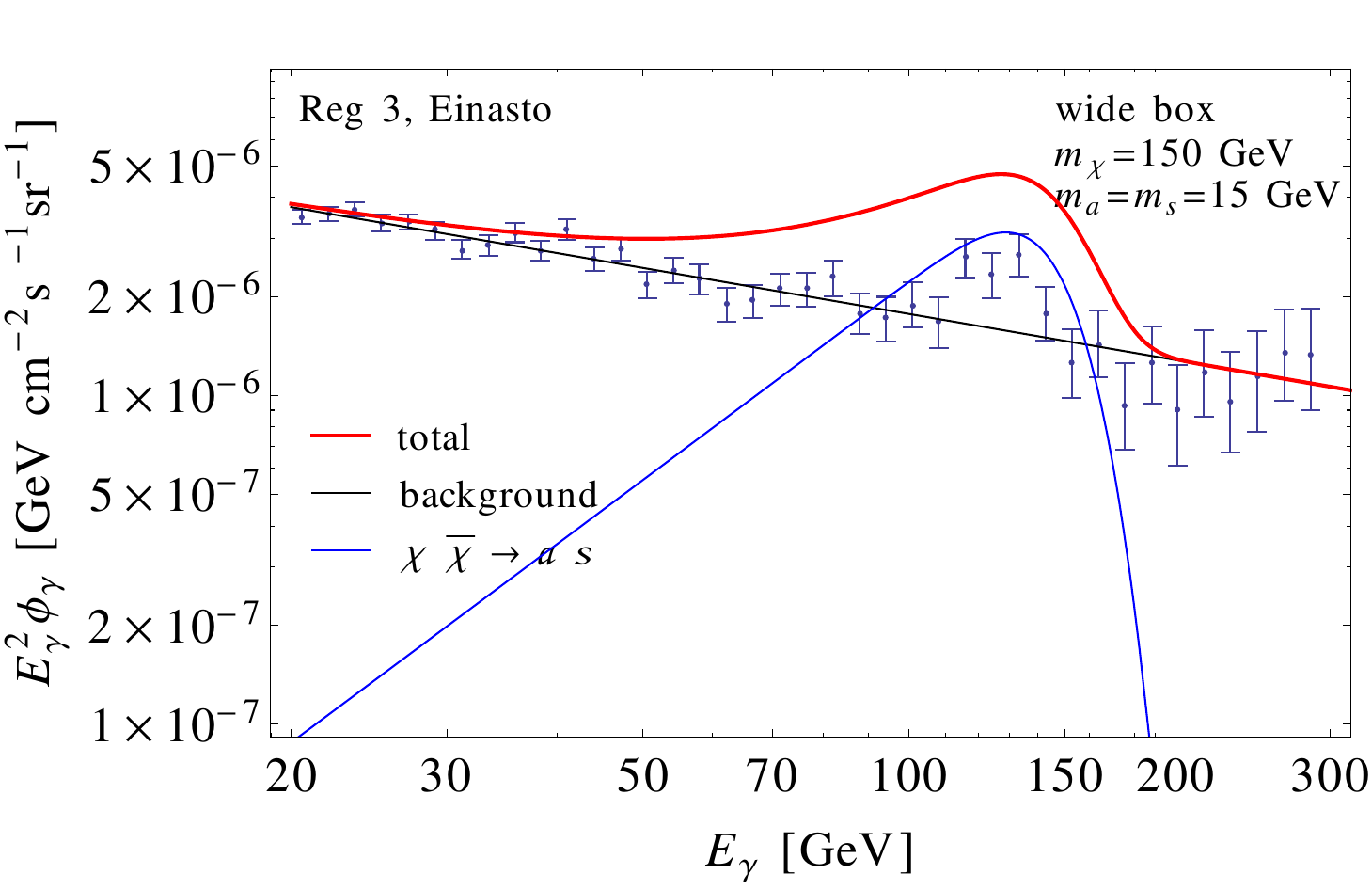}
\caption{\small Predicted gamma-ray spectrum for a narrow box scenario with $m_\chi=250\,{\rm GeV}, m_s=m_a=249.75\,{\rm GeV}$ assuming $c_1/c_2=3$ (left) and for a wide box scenario with $m_\chi=150\,{\rm GeV}$, $m_a=m_s=15\,{\rm GeV}$ (right). In both panels, the dark matter coupling $\lambda_\chi$ is determined as a function of dark matter mass by the thermal annihilation cross section $\langle\sigma v\rangle_{\rm th}=6\times 10^{-26}\,{\rm cm}^3 \,{\rm s}^{-1}$ (see Fig.~\ref{fig:lchi-vs-mchi}) and the present annihilation cross section follows from Fig.~\ref{fig:thermalrelic}. The branching ratio for the different annihilation channels and the branching ratios for the axion decay are predicted by the model, see text. The background flux is assumed to be a power law in the energy range shown and has been chosen to fit the data below $70\,{\rm GeV}$ (black line). 
}\label{fig:totalflux}
\end{figure}

 We consider first a narrow box scenario corresponding to  $m_\chi=250\,{\rm GeV}$, $m_s=m_a=0.999m_\chi=249.75\,{\rm GeV}$, shown as a star in Fig.~\ref{fig:thermalrelic}, and which gives ${\rm BR}(\chi\bar{\chi}\to as,aa,ss)\simeq 1,0,0$ and $\langle\sigma v\rangle_0/\langle\sigma v\rangle_{\rm fo}\simeq 0.238$. Furthermore, we consider the case with $c_1/c_2=3$, such that ${\rm BR}(a\rightarrow \gamma\gamma)\simeq 0.4$. The total flux predicted by the model for the region of the sky Reg3 defined in \cite{Weniger:2012tx} with this choice of parameters is shown in Fig.~\ref{fig:totalflux} (left) as a thick red line. Here, we adopted an Einasto profile and a background flux following the simple power law $\phi_\gamma^{\rm bg}=1.5\times 10^{-5} (E_\gamma/{\rm GeV})^{-2.46}\,{\rm GeV}^{-1}\,{\rm cm}^{-2}\,{\rm s}^{-1}\,{\rm sr}^{-1}$, which fits well the data below 70 GeV. The model predicts a gamma-ray feature that stands out over the power-law background at $\sim$130 GeV without invoking any boost factor. For other values of $c_1/c_2$ the gamma-ray feature can become fainter or stronger, however this change in intensity could be compensated by moderate variations of the astrophysical parameters, e.g.~a different choice of halo profile or an astrophysical boost factor that could enhance the signal by a factor of ${\cal O}(1)$. Note that for annihilations $\chi\bar{\chi}\to as$ a spectrum of continuum photons can be generated by the decays of $s$ and $a$, and therefore the corresponding constraints should be taken into account. These constraints on continuum photons typically place an upper limit of $\mathcal{O}(10-100)$ in the ratio of total cross section to line cross section \cite{Buchmuller:2012rc,Cohen:2012me,Cholis:2012fb}, so the models considered here are safe as long as the branching ratio of the process $a\rightarrow \gamma\gamma$ is sizeable.

\par The narrow box case analysed in the previous paragraph has fairly degenerate dark matter and axion masses and could arguably be considered fine-tuned. Nevertheless, the expectation of a gamma-ray excess holds in a much larger region of the parameter space.  We illustrate this fact using a second exemplary choice of parameters, corresponding to a wide box scenario and which in fact produces a gamma-ray flux overshooting the data. Concretely, we have taken $m_\chi=150\,{\rm GeV}$, $m_a=m_s=0.1m_\chi=15\,{\rm GeV}$, such that the axion decays dominantly into two photons, i.e.~${\rm BR}(a\to \gamma\gamma)\simeq 1$. This point is shown as a triangle in Fig.~\ref{fig:thermalrelic} and yields  ${\rm BR}(\chi\bar{\chi}\to as,aa,ss)\simeq 1,0,0$ and $\langle\sigma v\rangle_0/\langle\sigma v\rangle_{\rm fo}\simeq 1.013$. The total gamma-ray flux that results from this choice of parameters is shown in Fig.~\ref{fig:totalflux} (right) as a thick red line and turns out to be too intense to reproduce the Fermi-LAT observations. Again, moderate variations of the astrophysical parameters could lower the intensity of the signal producing a gamma-ray flux in better agreement with the data. However, and as we already mentioned above, we will not attempt to explain the 130 GeV excess within this framework. We merely present this example to illustrate that models producing gamma-ray boxes can generate, for the thermal cross section and typical astrophysical parameters, spectral features intense enough to be probed in present and future gamma-ray telescopes.

\section{Conclusions}
\label{sec:conclusions}

\par With several instruments providing observations of exquisite quality, it is now possible to search for fine spectral features in the gamma-ray emission from our Galaxy and beyond. This opportunity -- not present several years back -- is of extreme importance for dark matter searches and is perhaps one of the cleanest ways to glean on dark matter phenomenology. In the present contribution, we have focused on a particular type of spectral features, gamma-ray boxes, which stand out over the background at high energies. Interestingly, gamma-ray boxes are naturally realised in dark matter models without the need for fine-tuned parameters. Constructing a complete framework for axion-mediated dark matter, we show explicitly that both narrow and wide boxes are possible configurations of the photon spectrum induced by annihilations. Either way, thermal relics unambiguously foresee observable gamma-ray fluxes. On a more general note, we have also obtained model-independent constraints on box-shaped features from dark matter annihilation using the data provided by Fermi-LAT and H.E.S.S.. Our results -- that are, to the best of our knowledge, the first of their kind to extend to the multi-TeV range -- are remarkably strong despite our straightforward data analysis procedure. A dedicated box search would be an important step towards exploring the full potential of the current gamma-ray observations and would probably improve upon the constraints presented here. More importantly, judging from our findings, the prospects for future IACTs look very exciting given that such instruments will likely probe thermal cross sections up to several tens of TeV. The search for gamma-ray boxes in IACTs (in particular, CTA) rests thus as one of the last hopes to test thermal relics at high masses, a regime where other dark matter signals become increasingly dim, rendering direct detection, collider searches and other indirect detection channels of little relevance.

\section*{Acknowledgements} 
\par A.I.~and S.L.G.~would like to thank the Korea Institute for Advanced Study and the Sungkyunkwan University for hospitality during the first stages of this work and to Mathias Garny for useful discussions. This research was done in the context of the ERC Advanced Grant project ``FLAVOUR'' (267104) and was partially supported by the DFG cluster of excellence ``Origin and Structure of the Universe''. The work of H.~M.~L.~is supported in part by Basic Science Research Program through the National Research Foundation of Korea (NRF) funded by the Ministry of Education, Science and Technology (2013R1A1A2007919). M.~P.~acknowledges the support from Wenner-Gren Stiftelserna in Stockholm.


\begin{thebibliography}{999}

\bibitem{Bertone:2004pz}
  G.~Bertone, D.~Hooper and J.~Silk,
  Phys.\ Rept.\  {\bf 405} (2005) 279
  [hep-ph/0404175].

\bibitem{Bergstrom:2012fi}
  L.~Bergstrom,
  Annalen Phys.\  {\bf 524} (2012) 479
  [arXiv:1205.4882 [astro-ph.HE]].

\bibitem{Bertone:2010at}
  G.~Bertone,
  Nature {\bf 468} (2010) 389
  [arXiv:1011.3532 [astro-ph.CO]].

\bibitem{Srednicki:1985sf}
  M.~Srednicki, S.~Theisen and J.~Silk,
  Phys.\ Rev.\ Lett.\  {\bf 56} (1986) 263
   [Erratum-ibid.\  {\bf 56} (1986) 1883].

\bibitem{Rudaz:1986db}
  S.~Rudaz,
  Phys.\ Rev.\ Lett.\  {\bf 56} (1986) 2128.

\bibitem{Bergstrom:1988fp}
  L.~Bergstrom and H.~Snellman,
  Phys.\ Rev.\ D {\bf 37} (1988) 3737.

\bibitem{Bergstrom:1989jr}
  L.~Bergstrom,
  Phys.\ Lett.\ B {\bf 225} (1989) 372.

\bibitem{Flores:1989ru}
  R.~Flores, K.~A.~Olive and S.~Rudaz,
  Phys.\ Lett.\ B {\bf 232} (1989) 377.



\bibitem{Bringmann:2007nk}
  T.~Bringmann, L.~Bergstrom and J.~Edsjo,
  JHEP {\bf 0801} (2008) 049
  [arXiv:0710.3169 [hep-ph]].



\bibitem{Ibarra:2012dw}
  A.~Ibarra, S.~Lopez Gehler and M.~Pato,
  JCAP {\bf 1207} (2012) 043
  [arXiv:1205.0007 [hep-ph]].




\bibitem{Vertongen:2011mu}
  G.~Vertongen and C.~Weniger,
  JCAP {\bf 1105} (2011) 027
  [arXiv:1101.2610 [hep-ph]].


\bibitem{Ackermann:2012qk}
  M.~Ackermann {\it et al.}  [LAT Collaboration],
  Phys.\ Rev.\ D {\bf 86} (2012) 022002
  [arXiv:1205.2739 [astro-ph.HE]].



\bibitem{Abramowski:2013ax}
  A.~Abramowski {\it et al.}  [H.E.S.S. Collaboration],
  Phys.\ Rev.\ Lett.\  {\bf 110} (2013) 041301
  [arXiv:1301.1173 [astro-ph.HE]].

\bibitem{Abdo:2010dk}
  A.~A.~Abdo {\it et al.}  [Fermi-LAT Collaboration],
  JCAP {\bf 1004} (2010) 014
  [arXiv:1002.4415 [astro-ph.CO]].

\bibitem{Aleksic:2012cp}
  J.~Aleksic, J.~Rico and M.~Martinez,
  JCAP {\bf 1210} (2012) 032
  [arXiv:1209.5589 [astro-ph.HE]].


\bibitem{Bringmann:2012vr}
  T.~Bringmann, X.~Huang, A.~Ibarra, S.~Vogl and C.~Weniger,
  arXiv:1203.1312 [hep-ph].

\bibitem{Weniger:2012tx}
  C.~Weniger,
  arXiv:1204.2797 [hep-ph].

\bibitem{Su:2012ft}
  M.~Su and D.~P.~Finkbeiner,
  arXiv:1206.1616 [astro-ph.HE].

\bibitem{Bergstrom:2012vd}
  L.~Bergstrom, G.~Bertone, J.~Conrad, C.~Farnier and C.~Weniger,
  JCAP {\bf 1211} (2012) 025
  [arXiv:1207.6773 [hep-ph]].

\bibitem{Jungman:1994cg}
  G.~Jungman and M.~Kamionkowski,
  Phys.\ Rev.\ D {\bf 51} (1995) 3121
  [hep-ph/9501365].

\bibitem{Bergstrom:1997fh}
  L.~Bergstrom and P.~Ullio,
  Nucl.\ Phys.\ B {\bf 504} (1997) 27
  [hep-ph/9706232].

\bibitem{Bern:1997ng}
  Z.~Bern, P.~Gondolo and M.~Perelstein,
  Phys.\ Lett.\ B {\bf 411} (1997) 86
  [hep-ph/9706538].


\bibitem{Bergstrom:1997fj}
  L.~Bergstrom, P.~Ullio and J.~H.~Buckley,
  Astropart.\ Phys.\  {\bf 9} (1998) 137
  [astro-ph/9712318].

\bibitem{Gustafsson:2007pc}
  M.~Gustafsson, E.~Lundstrom, L.~Bergstrom and J.~Edsjo,
  Phys.\ Rev.\ Lett.\  {\bf 99} (2007) 041301
  [astro-ph/0703512 [ASTRO-PH]].

\bibitem{Bertone:2009cb}
  G.~Bertone, C.~B.~Jackson, G.~Shaughnessy, T.~M.~P.~Tait and A.~Vallinotto,
  Phys.\ Rev.\ D {\bf 80} (2009) 023512
  [arXiv:0904.1442 [astro-ph.HE]].

\bibitem{Dudas:2009uq}
  E.~Dudas, Y.~Mambrini, S.~Pokorski and A.~Romagnoni,
  JHEP {\bf 0908} (2009) 014
  [arXiv:0904.1745 [hep-ph]].

\bibitem{Mambrini:2009ad}
  Y.~Mambrini,
  JCAP {\bf 0912} (2009) 005
  [arXiv:0907.2918 [hep-ph]].

\bibitem{Jackson:2009kg}
  C.~B.~Jackson, G.~Servant, G.~Shaughnessy, T.~M.~P.~Tait and M.~Taoso,
  JCAP {\bf 1004} (2010) 004
  [arXiv:0912.0004 [hep-ph]].

\bibitem{Chu:2012qy}
  X.~Chu, T.~Hambye, T.~Scarna and M.~H.~G.~Tytgat,
  Phys.\ Rev.\ D {\bf 86} (2012) 083521
  [arXiv:1206.2279 [hep-ph]].



\bibitem{Bell:2008ey}
  N.~F.~Bell, J.~B.~Dent, T.~D.~Jacques and T.~J.~Weiler,
  Phys.\ Rev.\ D {\bf 78} (2008) 083540
  [arXiv:0805.3423 [hep-ph]].

\bibitem{Barger:2009xe}
  V.~Barger, Y.~Gao, W.~Y.~Keung and D.~Marfatia,
  Phys.\ Rev.\ D {\bf 80} (2009) 063537
  [arXiv:0906.3009 [hep-ph]].

\bibitem{Garny:2011cj}
  M.~Garny, A.~Ibarra and S.~Vogl,
  JCAP {\bf 1107} (2011) 028
  [arXiv:1105.5367 [hep-ph]].

\bibitem{Garny:2011ii}
  M.~Garny, A.~Ibarra and S.~Vogl,
  JCAP {\bf 1204} (2012) 033
  [arXiv:1112.5155 [hep-ph]].

\bibitem{Asano:2011ik}
  M.~Asano, T.~Bringmann and C.~Weniger,
  Phys.\ Lett.\ B {\bf 709} (2012) 128
  [arXiv:1112.5158 [hep-ph]].

\bibitem{Fan:2012gr}
  J.~Fan and M.~Reece,
  arXiv:1209.1097 [hep-ph].

  
\bibitem{axion1}
  H.~M.~Lee, M.~Park and W.~-I.~Park,
  Phys.\ Rev.\ D {\bf 86} (2012) 103502
  [arXiv:1205.4675 [hep-ph]];

\bibitem{axion15}
  H.~M.~Lee, M.~Park and W.~-I.~Park,
  JHEP {\bf 1212} (2012) 037
  [arXiv:1209.1955 [hep-ph]].

\bibitem{axion2}
  H.~M.~Lee, M.~Park and V.~Sanz,
  arXiv:1212.5647 [hep-ph].

\bibitem{Consortium:2010bc}
  M.~Actis {\it et al.}  [CTA Consortium Collaboration],
  Exper.\ Astron.\  {\bf 32} (2011) 193
  [arXiv:1008.3703 [astro-ph.IM]].


\bibitem{fermilatsite}
 http://www-glast.stanford.edu/.

\bibitem{hesssite}
 http://www.mpi-hd.mpg.de/hfm/HESS/.




\bibitem{Nomura:2008ru}
  Y.~Nomura and J.~Thaler,
  Phys.\ Rev.\ D {\bf 79} (2009) 075008
  [arXiv:0810.5397 [hep-ph]].

\bibitem{Mardon:2009gw}
  J.~Mardon, Y.~Nomura and J.~Thaler,
  Phys.\ Rev.\ D {\bf 80} (2009) 035013
  [arXiv:0905.3749 [hep-ph]].



\bibitem{Aharonian:2006au}
  F.~Aharonian {\it et al.}  [H.E.S.S. Collaboration],
  Nature {\bf 439} (2006) 695
  [astro-ph/0603021].


\bibitem{Rando:2009yq}
  R.~Rando [Fermi LAT Collaboration],
  arXiv:0907.0626 [astro-ph.IM].

\bibitem{fermilatsite2}
 http://www.slac.stanford.edu/exp/glast/groups/canda/lat\_Performance.htm.


\bibitem{Navarro:2008kc}
  J.~F.~Navarro, A.~Ludlow, V.~Springel, J.~Wang, M.~Vogelsberger, S.~D.~M.~White, A.~Jenkins and C.~S.~Frenk {\it et al.},
  arXiv:0810.1522 [astro-ph].


\bibitem{Hayashi:2007uk}
  E.~Hayashi and S.~D.~M.~White,
  arXiv:0709.3933 [astro-ph].


\bibitem{Gao:2007gh}
  L.~Gao, J.~F.~Navarro, S.~Cole, C.~Frenk, S.~D.~M.~White, V.~Springel, A.~Jenkins and A.~F.~Neto,
  arXiv:0711.0746 [astro-ph].




\bibitem{Navarro:2003ew}
  J.~F.~Navarro, E.~Hayashi, C.~Power, A.~Jenkins, C.~S.~Frenk, S.~D.~M.~White, V.~Springel and J.~Stadel {\it et al.},
  Mon.\ Not.\ Roy.\ Astron.\ Soc.\  {\bf 349} (2004) 1039
  [astro-ph/0311231].

\bibitem{Springel:2008cc}
  V.~Springel, J.~Wang, M.~Vogelsberger, A.~Ludlow, A.~Jenkins, A.~Helmi, J.~F.~Navarro and C.~S.~Frenk {\it et al.},
  Mon.\ Not.\ Roy.\ Astron.\ Soc.\  {\bf 391} (2008) 1685
  [arXiv:0809.0898 [astro-ph]].

\bibitem{Catena:2009mf}
  R.~Catena and P.~Ullio,
  JCAP {\bf 1008} (2010) 004
  [arXiv:0907.0018 [astro-ph.CO]].

\bibitem{Weber:2009pt}
  M.~Weber and W.~de Boer,
  Astron.\ Astrophys.\  {\bf 509} (2010) A25
  [arXiv:0910.4272 [astro-ph.CO]].

\bibitem{Salucci:2010qr}
  P.~Salucci, F.~Nesti, G.~Gentile and C.~F.~Martins,
  Astron.\ Astrophys.\  {\bf 523} (2010) A83
  [arXiv:1003.3101 [astro-ph.GA]].

\bibitem{Pato:2010yq}
  M.~Pato, O.~Agertz, G.~Bertone, B.~Moore and R.~Teyssier,
  Phys.\ Rev.\ D {\bf 82} (2010) 023531
  [arXiv:1006.1322 [astro-ph.HE]].


\bibitem{Buchmuller:2012rc}
  W.~Buchmuller and M.~Garny,
  JCAP {\bf 1208} (2012) 035
  [arXiv:1206.7056 [hep-ph]].


\bibitem{Cohen:2012me}
  T.~Cohen, M.~Lisanti, T.~R.~Slatyer and J.~G.~Wacker,
  JHEP {\bf 1210} (2012) 134
  [arXiv:1207.0800 [hep-ph]].

\bibitem{Cholis:2012fb}
  I.~Cholis, M.~Tavakoli and P.~Ullio,
  Phys.\ Rev.\ D {\bf 86} (2012) 083525
  [arXiv:1207.1468 [hep-ph]].

\bibitem{Gondolo:1990dk}
  P.~Gondolo and G.~Gelmini,
  Nucl.\ Phys.\ B {\bf 360} (1991) 145.
  doi:10.1016/0550-3213(91)90438-4




\end{thebibliography}
\end{document}